# Spreadsheets in Financial Departments:

## An Automated Analysis of 65,000 Spreadsheets using the *Luminous* Technology

Dr. Kevin McDaid, Dr Ronan MacRuairi, Mr. Neil Clynch, Mr. Kevin Logue, Mr. Cian Clancy, Mr. Shane Hayes

Spreadsheet Research Group

Dundalk Institute of Technology, Dundalk, Co. Louth, ROI,

kevin.mcdaid@dkit.ie

**ABSTRACT**

*Spreadsheet technology is a cornerstone of IT systems in most organisations. It is often the glue that binds more structured transaction-based systems together. Financial operations are a case in point where spreadsheets fill the gaps left by dedicated accounting systems, particularly covering reporting and business process operations.*

*However, little is understood as to the nature of spreadsheet usage in organisations and the contents and structure of these spreadsheets as they relate to key business functions with few, if any, comprehensive analyses of spreadsheet repositories in real organisations.*

*As such this paper represents an important attempt at profiling real and substantial spreadsheet repositories. Using the **Luminous** technology an analysis of 65,000 spreadsheets for the financial departments of both a government and a private commercial organisation was conducted. This provides an important insight into the nature and structure of these spreadsheets, the links between them, the existence and nature of macros and the level of repetitive processes performed through the spreadsheets.*

*Furthermore it highlights the organisational dependence on spreadsheets and the range and number of spreadsheets dealt with by individuals on a daily basis. In so doing, this paper prompts important questions that can frame future research in the domain.*

## 1. INTRODUCTION

Despite the low level of research in the domain, spreadsheet technology is as fundamental to the smooth running of every financial department as the dedicated financial accounting system that may be in place.

However, with its poor level of control and visibility, spreadsheet usage often slips beneath the radar of management. As a result, an organisation is rarely aware of its dependence on spreadsheets, and of the nature of these spreadsheets and their relevance to key business processes. Furthermore they are rarely aware of the usage of spreadsheets by individual employees and whether this usage is appropriate to the level and experience of the employee.

Dundalk Institute of Technology, along with **CUSPA Technologies**, has developed a suite of technologies, termed **Luminous**, which includes a tool for rapidly scanning and analysing large repositories of spreadsheets to discover an organisation's dependence on spreadsheet usage for key business processes. The tool has been successfully deployed at a number of organisations including one at which in excess of a quarter of a million spreadsheets were analysed containing 4 billion cells of information.

This paper outlines the results of the tool's application to the spreadsheet repositories of the financial departments of two firms, one a government sector company of approximately 400 employees, and the second a private commercial company with annual turnover in excess of 300 million euro. In total more than 65,000 spreadsheets were analysed.For the purposes of confidentiality, the names of companies, individuals, and spreadsheet files have been altered.

The paper profiles the spreadsheets in terms of size, cell content and use of functions and other features. It summarises the use of macros and measures the extent of external links. Crucially it examines the use of spreadsheets across business processes over time. Finally, it highlights numerous poor practices involving the linking of spreadsheets.

To the best of our knowledge this paper represents one of the first attempts to comprehensively profile real spreadsheet usage through direct analysis of complete real and current repositories. As such it represents important work in the domain of spreadsheet usage.

The remainder of this paper is structured as follows: Section 2 provides the background for this analysis.  Section 3 presents the technology used for collating and analysing data. Section 4 presents results of the respective repository scans. Section 5 discusses the work and draws conclusions, and Section 6concludes this paper.

## 2. BACKGROUND

This paper analyses the centrally stored spreadsheet repositories associated with the activities of the financial departments of two large firms.

The first firm, operating in the government sector, has in excess of 400 employees with approximately 10 people employed in financial management and administrative roles. For this paper we name this company *Company A*. The second firm operates in the Wholesale Food sector employing 1700 people with an annual turnover of approximately 350 million euro. The financial function is larger, with up to 20 people. For this paper we name this company *Company B*.

As financial departments, the activities of the individuals of the firms will be similar, dealing with core financial and accounting systems. The main difference is that Company B would be involved in significant sales activity, whereas Company A would not. As with most firms spreadsheets are used in many instances act as the glue between the formal systems that allows their smooth, or not so smooth, integration, and in other instances as the tool of choice for reporting. Finally, spreadsheets are also used to automate self contained business processes, separate from the more structured and controlled systems.

The analysis is not conducted on all servers and PC's for the firms, instead it is pointed only at the key shared repositories. These represent the primary storage areas for the organisations. For Company A, the repository is located on a server running Novell Netware 6.5 with machines accessing this repository typically running Windows XP, and using Microsoft Office 2003. For Company B, the repository is located on a Windows 2003 Server.  Machines accessing this repository are running either a version of Windows 2000or Windows XP and using Microsoft Office 2003 or Office 2007.

The*Luminous*application found 12,378 spreadsheet files for Company A and 57,446 files for Company B.

**Related Research**

We understand that a number of vendors of Enterprise Spreadsheet Management software, including CIMCON at Conference of European Spreadsheet Research Interest Group 2010, have presented summary information of server based repositories.However, we are not aware of any other formally published research that profiles entire spreadsheet repositories, particularly with regard to the business function of the organisation.

That said, numerous studies have examined real spreadsheets. In particular Powell, Baker and Lawson have examined for correctness a very small sample of spreadsheets from 5 different firms [1]. The spreadsheets were selected by the firm and this cannot be considered as a random sample from the organisations. Other authors have also examined real spreadsheets for the purposes of determining reliability. This work is summarised by Panko in [2] with a number of repositories of spreadsheet errors presented by the same author in [3].The most complete survey of spreadsheet usage based on survey methods is given in [4], also by Powell, Baker and Lawson.

The EUSES consortium has gathered a repository of approximately 4000 spreadsheets for research purposes. They have used this set in a number of publications including [5]. The question arises as to how well this group reflects use of spreadsheets in industry. We investigate this repository and compare it with the spreadsheets we have analysed.

Finally, Chambers and Hamill [6] and McGeady and McGouran [7] examine spreadsheet use across financial organisations in the context of how best to control spreadsheets in organisations.

### 3. TECHNOLOGY

The*Luminous* toolis a standalone application that can analyse large repositories of Microsoft Excel spreadsheets in a highly efficient manner. Importantly, it does not require an installation of Microsoft Office on the host machine to run and can be pointed at a single or a number of mapped server and/or local locations. Its efficiency has been demonstrated by the successful application to other repositories of in excess of quarter of a million spreadsheets covering in excess of 100Gb of spreadsheet information.

*Luminous* analyses spreadsheets on a file by file basis. It does notprocess files of type older than Microsoft Excel 95, and files that are password protected can only be analysed if passwords are provided. For the analyses of Company A and Company B passwords were not provided and so password protected files were not analysed.

*Luminous*provides a wealth of information on each spreadsheet including file level information such as size, user who last saved the spreadsheet, date file created, date last modified and location.

In addition, based on an individual analysis of each cell and object in each spreadsheet the application provides details of cell content, use of inbuilt Excel function usage, existence of objects such as pivot tables etc, extent and nature of links between and across spreadsheet files. Furthermore, Luminous providesan add-on to the tool allows the analysis of macros contained in spreadsheets.

### 4. RESULTS

The analyses of the repositories are summarised in this section. Given the extent of information gathered through the application and the size of the repositories, this paper represents an initial summary rather than a complete analysis of the data. Following further analysis we intend to publish more detailed results. Data for Company A is first presented followed by data for Company B. There is some analysis included for Company B that is not, due to pressure of space, included for Company A.

### 4.1 Company A

*File Size*

**Luminous** detected 12,378 workbooks, successfully processing 12,161 (98%), amounting to 11Gb of data. Of the remaining 217 workbooks, 157 are pre-1995 Excel files not supported by the scan technology; 12 are password protected; and the remaining 48 were notprocessed due to being either corrupt or of an older format. Table 1 below shows a distribution of file sizes. Note that the largest file was 280Mb, and that 13% of files are greater than 1Mb in size.

Table 1: Distribution of File Size

| File Size | <10Kb | 10Kb - 50Kb | 50Kb - 100Kb | 100Kb - 500Kb | 500Kb - 1 Mb | 1Mb - 10 Mb | 10Mb - 50Mb | 50Mb - 100Mb | 100Mb - 150Mb | 150Mb - 200Mb | >200Mb |
|---|---|---|---|---|---|---|---|---|---|---|---|
| Number of Files | 75 | 5,122 | 1,956 | 2,839 | 557 | 1,520 | 62 | 10 | 2 | 17 | 1 |
| Percentage | 1% | 42% | 16% | 23% | 5% | 12% | 1% | 0% | 0% | 0% | 0% |

*Structure of Spreadsheets*

The analysis of individual worksheets indicates that on average a file contains approximately 160 used columns and 1,500 used rows spread over 7 sheets involving approximately 10,000 cells on average. However the median values are lower with 3 sheets, 185 rows and 23 columns illustrating the skewed nature of the distribution for this data.

The breakdown by file of the number of rows, columns and sheets across the repository is presented in Table 2 below. It shows the number of spreadsheets which contain more than the stated number of rows, columns or sheets. Interestingly it shows that 18% spreadsheets contain more than 10 worksheets. In this context we will later examine how workbooks with a large number of sheets are used to manage repeated business process activities over time.

Table 2: Distribution of Percentage of Rows, Columns and Sheets for Company A

| Criteria | <=10 | >10 | >50 | >100 | >250 | >500 | >750 | >1,000 | >10,000 |
|---|---|---|---|---|---|---|---|---|---|
| Worksheets | 82% | 18% | 1% | 0% | 0% | 0% | 0% | 0% | 0% |
| Columns | 21% | 79% | 32% | 20% | 9% | 4% | 1% | 1% | 0% |
| Rows | 2% | 98% | 79% | 64% | 43% | 32% | 24% | 20% | 3% |

It is interesting to examine the content of the workbooks in that we found that of the 119 million cells containing information 43% contained text data, 25% contained numeric data and 32% contained formulas.

*Use of Inbuilt functions*

Inbuilt functions in Microsoft Excel are a powerful feature of the spreadsheet application. Our analysis showed that there was no use of the following functions: INDIRECT, TRANSPOSE, INDEX, MATCH, ISERROR, MMULT and AVERAGEIF; and that the IRR function was used in just one cell in the entire repository. Table 3 presents information on the number of cells that reference each of the usedfunctions.

Table 3: Occurrence of Functions for Company A

| Criteria | OFFSET | VLOOKUP | HLOOKUP | NPV | SUM |
|---|---|---|---|---|---|
| Percentage of files with function | 0.01% | 11.41% | 1.23% | 0.02% | 65.90% |
| Max number of functions in individual file | 35 | 229,638 | 18,584 | 2 | 29,317 |
| Average number per file when present | 35 | 14,698 | 9,617 | 2 | 723 |

| Criteria | SUMIF | AVERAGE | COUNT | COUNTIF | IF |
|---|---|---|---|---|---|
| Percentage of files with function | 0.26% | 0.26% | 1.17% | 0.09% | 14.41% |
| Max number of functions in individual file | 763 | 42 | 18 | 414 | 114,841 |
| Average number per file when present | 474 | 12 | 1 | 264 | 6,430 |

The data shows that, for example, 0.09% of the spreadsheets use the COUNTIF function with an average of 264 cells in these workbooks using the function. One file included 414 cells that call the function.

The most popular functions are SUM, IF and VLOOKUP. Surprisingly, while VLOOKUP is called in 11% of all files compared with 66% for the SUM function, the average number of times the VLOOKUP function is used in files which contain the function is 14,698 compared with 723 for the SUM function.

Importantly, on examination of some of the files with a large number of VLOOKUP calls, we found that these workbooks often consisted of large sets of disparate data extracted from applications such as accounting systems and involving the matching of these data sets to standard lists. The question must arise as to the efficiency of these operations and the potential for investment in automated solutions, either embedded in Excel or not, to reduce the time and to increase the reliability of these tasks.

The application detected that 2.7% of files contained pivot tables, withanaverage 2 present per file for those files with pivot tables, and that 1% of files contain active auto filters with an average of 4 in use. The application also detected 570 workbooks (4%) that contain a macro of some sort. More detailed macro analysis is presented later for Company B.

*External Links*

Spreadsheets can be linked together through formulas that refer to other spreadsheets. In this way chains of spreadsheets can share information to perform complex business processes. For Company A we found that 26.0% (3,162) of all spreadsheets contained at least one external link. For those with external links we found that each spreadsheet contains on average 4,270 cells that refer to other spreadsheets. More interestingly, although the average number of external workbooks referred to is 2, there are 85 workbooks that refer to more than 10 external workbooks. Additionally, 17 spreadsheets contained in excess of 100,000 cells that reference other external workbooks.

As an indication of the internal structure we found that 25.8% of spreadsheets contain worksheets that link to other worksheets in the workbook with an average of 1,266 cells that link to other worksheets in the same workbook.

**4.2 Company B**

*Luminous* detected 57,446 workbooks and successfully processed 53,645 workbooks (93%), amounting to 40Gb of data. Of the remaining 3,801 workbooks 536 are pre-1995 Excel files with 3,263 files password protected, and the remaining 2 were notprocessed due to being corrupt.

*Filesize*

Table 4 shows the distribution of file size for Company B. The largest file is 105Mb, and there are 19% of files greater than 1Mb in size, higher than the 14% quoted for Company A. For both companies the percentage of files below 50kb is just above 40%. Note that the median filesize was 85kb.

Table 4: Distribution of File Sizes for Company B

| File Size | <10Kb | 10Kb - 50Kb | 50Kb - 100Kb | 100Kb - 500Kb | 500Kb - 1 Mb | 1Mb - 10 Mb | 10Mb - 50Mb | 50Mb - 100Mb | 100Mb - 150Mb | 150Mb - 200Mb | >200Mb |
|---|---|---|---|---|---|---|---|---|---|---|---|
| Number of Files | 26 | 21,549 | 6,921 | 11,899 | 2,931 | 9,755 | 550 | 13 | 1 | 0 | 0 |
| Percentage | 0.1% | 40.2% | 12.9% | 22.2% | 5.5% | 18.2% | 1.03% | 0.02% | 0.00% | 0.00% | 0.00% |

*Structure of Spreadsheets*

As with Company A the average number of sheets in a workbook is approximately 7. For Company B, the average number of rows used was higher, at 2,300 compared to 1,500 (for Company A); however, Company B had a lower number of columns used – 90 compared to 150. The distribution of values for Company B is shown in Table 5 below.

Table 5: Distribution of Number of Rows,Columns and Sheets for Company B

| Criteria | <=10 | >10 | >50 | >100 | >250 | >500 | >750 | >1000 | >10000 |
|---|---|---|---|---|---|---|---|---|---|
| Worksheets | 87.3% | 12.7% | 1.7% | 0.8% | 0.2% | 0.0% | 0.0% | 0.0% | 0.0% |
| Columns | 24% | 76% | 31% | 19% | 9% | 3% | 2% | 1% | 0% |
| Rows | 2% | 98% | 77% | 63% | 45% | 33% | 26% | 22% | 4% |

Thirteen per cent of spreadsheets contained in excess of 10 worksheets, less than the 18% for Company A. On the other hand there were 82(0.2%) workbooks each containing more than 250 worksheets. Company A contained no such workbooks.

The average number of used cells was higher, at 17,000 approximately with 40% of cells containing text, 37% of cells containing numeric data and 23% containing formulas based on an analysis of almost 900 million cells of information.

*Use of inbuilt functions*

For Company B there was no use of the following functions: TRANSPOSE, MATCH, MMULT, AVERAGEIF, NPV and IRR.

As with Company A, the most used functions were SUM, VLOOKUP and IF with 71% of files containing at least one SUM function, 14% containing at least one IF function and 6% containing at least one VLOOKUP. Again, the use of VLOOKUP was extensive for the files in which it was used. The use of the IF function is also very high with almost 25,000 cells using it for files where it is present. As with Company A, the use of other functions in Company B was minimal, apart from the ISERROR function which occurs in almost 3% of files.

Finally, the use of pivot tables was higher at 4% with the use of auto filters significantly higher at nearly 5% of workbooks. More detailed results for all functions and pivottables and autofilters is shown in Table 6.

Table 6: Number of Occurrences of Functions for Company B

| Criteria | OFFSET | VLOOKUP | HLOOKUP | NPV | SUM |
|---|---|---|---|---|---|
| Percentage of files with function | 0.86% | 6.45% | 0.48% | 0.00% | 70.50% |
| Max number of functions in individual file | 9,958 | 284,896 | 42,744 | 0 | 69,349 |
| Average number per file when present | 150 | 4,862 | 825 | | 456 |
| | | | | | |
| Criteria | SUMIF | AVERAGE | COUNT | COUNTIF | IF |
| Percentage of files with function | 2.62% | 1.62% | 1.70% | 0.64% | 14.38% |
| Max number of functions in individual file | 8,076 | 2,977 | 30,702 | 3,648 | 845,928 |
| Average number per file when present | 882 | 215 | 43 | 211 | 24,842 |
| | | | | | |
| Criteria | INDIRECT | INDEX | ISERROR | | |
| Percentage of files with function | 0.02% | 0.14% | 2.81% | | |
| Max number of functions in individual file | 12,584 | 148 | 59,845 | | |
| Average number per file when present | 1,572 | 138 | 2,898 | | |

*External Links*

For Company B, the percentage of files with links to external workbooks was lower, but still significant. 17% of spreadsheets access data from a separate workbook, with each of these workbooks accessing three other workbooks on average. In terms of internal structure, we found that for both Company A and Company B, 26% of spreadsheets contain worksheets that link to other worksheets in the workbook. However, Company B has a higher average of 1,630 cells that link to other worksheets in the same workbook.

The question of the extent and integrity of external links is an important one, and **Luminous** technology helps to answer this via production of network maps to explain the links between files. We show below one real example (names of files have been minimally altered) where the file Summary Update.xls accesses data from 7 other files. However, there are issues with two of those files. In the first case Copy of Repair and Rental.xls does not exist in the location being called and, in the second case, Food Rec.xls exists but the specific sheet being called does not. We have labelled this as a ghost sheet in the diagram.

We have found that this is not an isolated case and that many of the external links for spreadsheets are effectively broken. Furthermore, there are many cases of files on the server linking to files on local machines. While it would be natural for spreadsheets on local machines to draw data from a central spreadsheet located on the server, it is certainly important to monitor and ensure the integrity of links going in the opposite direction.

In addition we found that many links exist to files that reside in temporary storage spaces on the local machines, many of which indicate links to spreadsheets that were attached to emails and opened from the associated temporary storage locations. This practice indicates a clear lack of awareness of file management.

We acknowledge that many of the links may be legacy ones that may well not have a material impact on the correctness of the spreadsheet for its current use, but the research indicates that this is an area that merits further attention. To that end we intend to conduct more detailed analysis in this area for future publication.

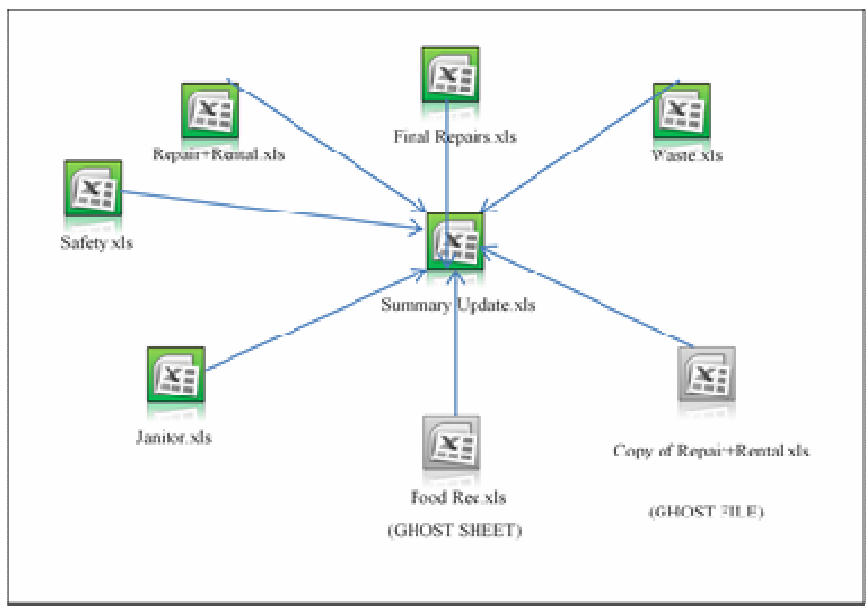

Figure 1: Link Map with Ghost File and Ghost Sheet

There is also a clear issue with the complexity of the link maps that can arise. We present in Figure 2 below a second real example.In this example, *Day Bank Report.xls* has 9 immediate source files, and is a precedent for over 250 other workbooks! One of these dependents, *Daily FX.xls*, has 45 dependents, one of which, *Treasury Dec.xls*, has an invalid data source.

While spreadsheets can be linked to create, what could be termed spreadsheet information systems, the management of such a network will be very problematic, particularly if more than one user is responsible for the different files.

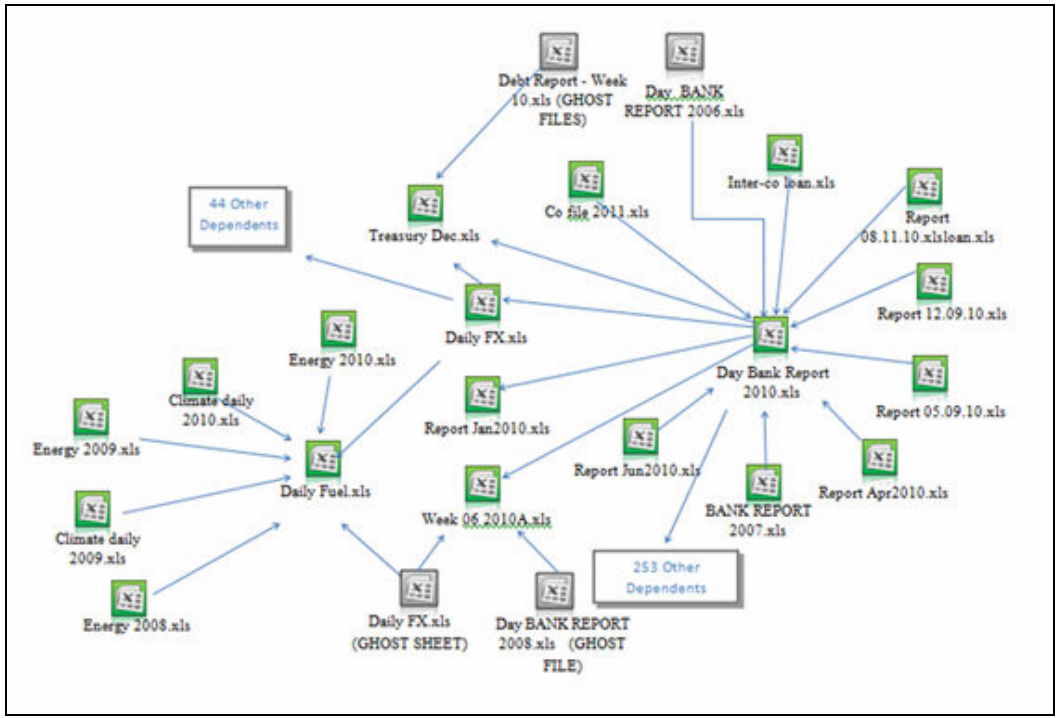

Figure 2: Complex Link Map

*Macro analysis*

We include a brief analysis of the macros in the workbooks for Company B. Of the 53,645 workbooks 1,750 workbooks (3%) contain a macro with 1,734 analysed in detail. The average number of lines of code per macro was 56, with 134 containing over 250 lines and 84 workbooks containing over 500 lines of macro code.

*Spreadsheets by Date and User*

It is interesting to examine the creation and saving of spreadsheets by time and user. For company B the date of last modification for the files starts at 1997 through to March 2011. We show in Figure 3 below a plot of the number of files by date created and by date last modified for months since January 2010. The graph shows that between 600 and 1000 files are created and last modified each month.

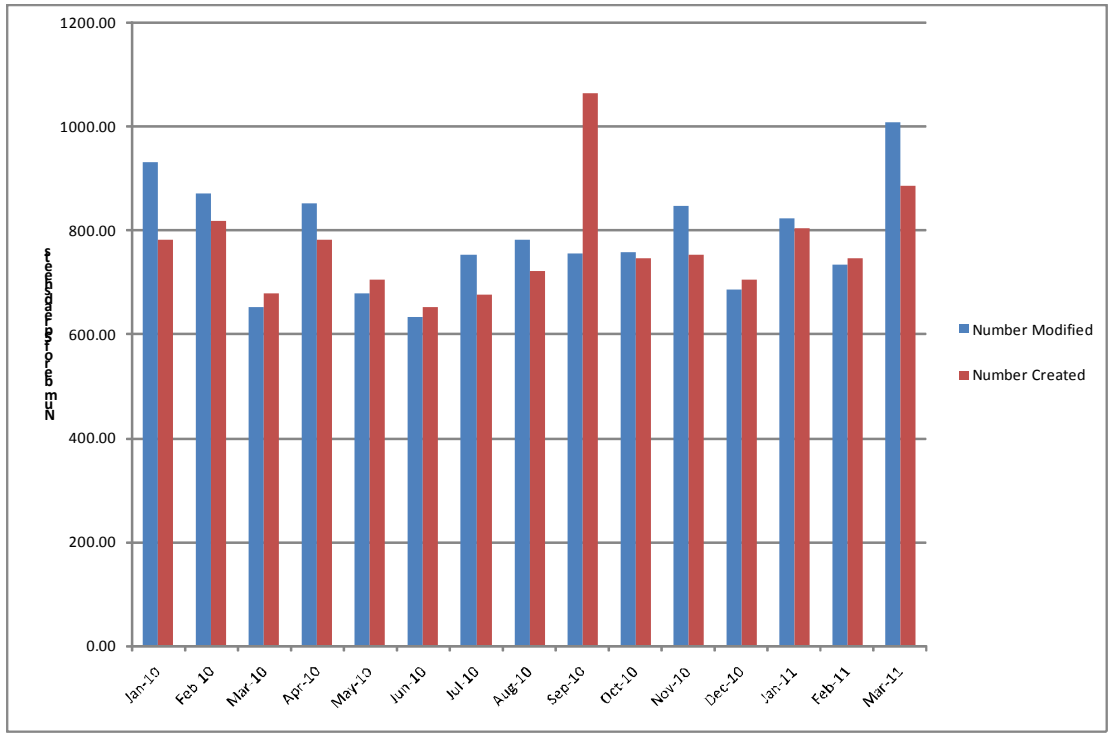

Figure 3: Number of Files by Date Created and Date Last Modified

Concentrating on the files that have been modified since January 2010 we plot the number last saved by each user, showing the number foreach user. The plot ignores those files automatically generated through export from accounting and other systems. In total there are66 different users in the data set,with only 43 users with more than 10 spreadsheets. Interestingly, the 8 heaviest spreadsheet users are each the last saved user for more than 400 spreadsheets over the 15 month period.

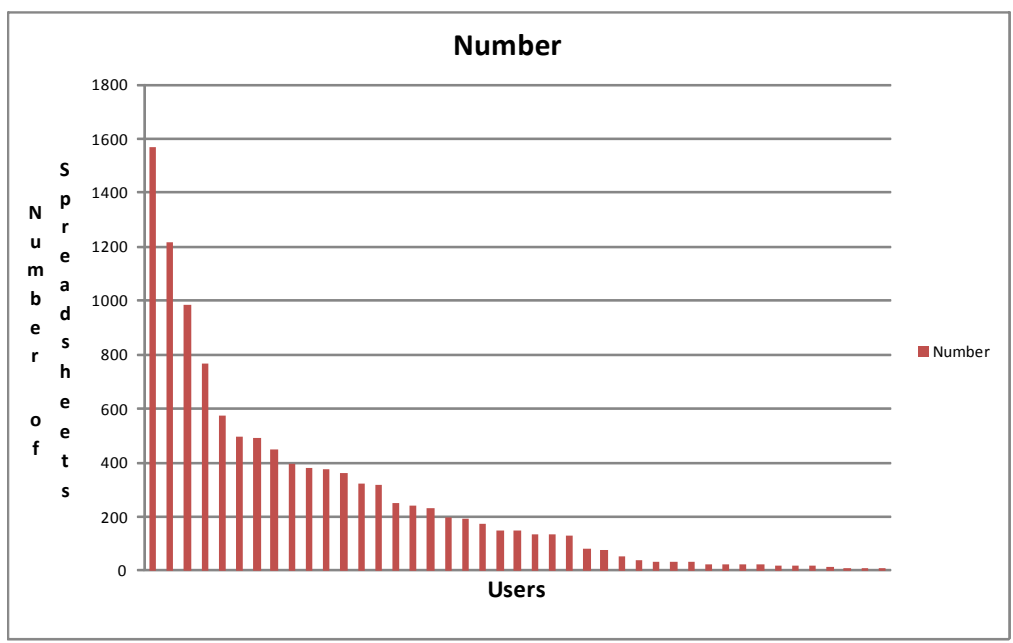

Figure 4: Number of Spreadsheets last saved by each user

*Spreadsheet Use for Repeated Business Processes*

Spreadsheets can be used for single once off tasks. They can also be used for business activities that repeat every day, month, week or quarter or even year. On closer examination we found that the use of spreadsheets for repeated activities can be achieved through the creation of a new workbook for each period or through the creation of a new sheet for each period.

Through examination the folder structure and file names, which were very detailed and well structured, we established that the level of use of spreadsheets for repeated activities was highly significant particularly for monthly, weekly and even daily accounting tasks.This provides a key insight into the issue of spreadsheet proliferation.

We also found, through examination of sheet names, that many activities are repeated through creation of an additional sheet for each period. We found one example of a workbook with 356 sheets, each representing a week from late 2003 through to 2011.

## 5. DISCUSSION & FINDINGS

This work is very much investigative in nature. At the outset we did not embark on the work with any prior hypotheses as to extent and nature of the repositories we would be analysing. The work seeks to provide an overview, using empirical methods, of the spreadsheets. The work does not explore the usage and development processes for the spreadsheets. While there were discussions with the Chief Financial Officer and with the Head of Finance of Company B, there were no detailed exchanges on the content and function of the spreadsheets within business processes. However, we presented these results to the senior staff,and they were very much in agreement with our findings, and further work is ongoing with Company B to seek to improve the organisation's use of spreadsheets within their processes.

There is much further analysis of the data that is possible, particularly with regard to the extent of links that are obsolete or are linked to temporary internet files. It is also planned that we will examine the information to establish the any relationships that may exist between key factors. In addition, there are many ways in which the tool might be improved. One interesting avenue would involve the analysis of cell formulas to separate cells that perform computations from ones that merely relay information contained in other cells possible originating in different worksheets or workbooks. Another improvement would be to measure the number of unique or root formulas that includes each of the functions as well as the number of times these unique formulas are repeated.

A summary of our key findings and the lessons for the research and practitioner community follows. In some cases these findings are unsurprising, but in others they may provide the stimulus for important areas of future research. In considering these findings, we acknowledge the threats to the external validity of these results based on the fact that the results are for just two departments in a single country in two entirely different economic sectors. We would argue that the users of the spreadsheets are very much non-experts, with a very small number of more expert users. However, no formal profiling was conducted.

We also acknowledge that the analysis presented here is very much a high level view. However, and most importantly, we believe that this work represents a natural starting point for the analysis of spreadsheet use in organisations and that the work is important given the lack, to our knowledge, of similar detailed work

Our key findings are:

- In these organisations the use of excel functions is largely restricted to SUM, IF and VLOOKUP. However the use of VLOOKUP is, to us, surprisingly large with many spreadsheets including functionality, implemented through VLOOKUP, to match very large lists often extracted automatically from accounting systems. Discussions with managers indicated that the development of formulas with VLOOKUP is not due to all users, but rather to a smaller number of more advanced users. However, these spreadsheets are often shared, and given the complexity and issues with VLOOKUP, this may represent a significant risk.

- The prevalence of the IF function is also significance. This is not surprising but to us the extent of its use, on average nearly 25,000 times for Company B, is.

- Although the users are not advanced, there are significant links between workbooks, and the integrity of these links is a clear issue, given that many are calling local files and many other calling files in temporary folders. When presented with link diagrams, managers expressed concern at some of the links which were to older versions of files of ones from previous years and seemed to be unaware of the impact of renaming of sheets and workbooks. The extent of the chains of spreadsheets that result surprised us and, given that many users are unaware of managing links, there are significant issues with the integrity of the links. We acknowledge that many of these links may be legacy ones without any impact on the materiality of the spreadsheet.

- The organisations use spreadsheets extensively for repeating business activities such as weekly, monthly and even, in the case of Company B, daily accounts. This is often performed through copying of the workbook or worksheet for the preceding period and the replacement with up to date values. There are clearly efficiency issues with these activities, and the question for the organisation must be whether a structured automated solution might be better. Alternatively, there may well be an argument to the development of templates or solutions developed through VBA technology.

- Finally, while the majority of spreadsheets are relatively small in size, the organisations develop a high number of large and possibly unmanageable spreadsheets, often arising from the repeated business activities as highlighted in the previous bullet point.

In our view a number of specific lessons can be learned from this research:

1. Researchers and practitioners need to address the extensive use of VLOOKUP in spreadsheets to both establish its reliability and ensure its efficient and effective usage. The work raises the need for technologies that serve to combine, integrate and consolidate large datasets in spreadsheets. In that regard the PowerPivot functionality in Excel 2010 may reduce somewhat the dependence on the VLOOKUP function.

2. There is a need for better knowledge, visibility and management of links between spreadsheet workbooks. There may well be commercial technologies that help in this regard but the issue in the first instance is one of awareness and understanding.

3. There are efficiency issues with the repetition of tasks in spreadsheets, and organisations seem to be unaware of the potential of alternative solutions or use of advanced features of spreadsheets to improve matters.

4. Discussions with managers on the use of VLOOKUP and IF functions highlighted the disparity in knowledge among users of spreadsheets across the organisations. They also highlighted the sharing of workbooks containing elements that not all users were entirely familiar with. This prompts the need for focussed training to counter this disparity and for structuring and protection of workbooks to reduce the associated risks.

More generally, this research prompts the question as to whether the focus of spreadsheet research fits the profile of spreadsheet development in industry as highlighted here. This research shows that many of the spreadsheets are large with limited use of complex functions. There are real efficiency issues around the development and use of this type of spreadsheet. To date, possibly for practical reasons, most of the controlled experiments conducted by researchers such as Panko[2] and Bishop and McDaid[8] deal with smaller spreadsheets.

To guide future research it is important to have access to real and entire spreadsheet repositories. The only publicly available repository is the EUSES one which represents a disparate collection with a large number produced by students for spreadsheet assignments. There is a question as to how representative they are of spreadsheet use in general. We applied the *Luminous* technology to this set of spreadsheets with interesting results. While we do not have space to detail all the results we found that in general, the EUSES spreadsheets are significantly smaller than the organisations' ones analysed with on average 31 columns and 296 rows. There was a much lower use of VLOOKUP, SUM and IF with less than 0.2% of files containing VLOOKUP, and, given the disparate sources, there was a very low level of external linking.

This certainly highlights the need for commercial practitioners and organisations to make available spreadsheets for research purposes.

## 6. CONCLUSION

The use of spreadsheets in real firms is certainly well understood by many expert practitioners who make a living in this domain. However, to the best of our knowledge this work represents the first detailed analysis of complete spreadsheet repositories for two finance departments.

Using the highly efficient *Luminous* technology we analysed in excess of 65,000 spreadsheets. The analysis was very much empirical in nature concentrating on the structure, functional usage, linking, macro content as well as user and temporal file information.

The analysis found that while many of the spreadsheets were large, they contained a low level of use of in built functions outside of the SUM function. However, there was evidence of relatively high usage of the VLOOKUP and IF functions, extensive linking of spreadsheets and the widespread repetitive use of spreadsheets for periodic business activities. The paper highlights the disparity in spreadsheet knowledge within firms and the lack of knowledge of advanced features. It also highlights potential deficiencies with publicly available spreadsheet repositories and calls for practitioners and organisations to make real spreadsheets available for research purposes.

## ACKNOWLEDGEMENT

This work was supported by Enterprise Ireland through Project CFTD/2009/0216.

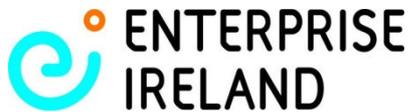
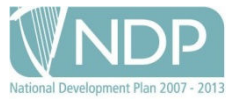
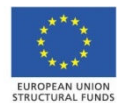